\documentclass{ifacconf}

\usepackage{amsmath,amssymb,amsfonts}
\usepackage{pifont}
\usepackage{enumitem}
\usepackage{graphicx}
\usepackage{cuted}
\usepackage{color}
\usepackage{flushend}
\usepackage{mathtools}
\usepackage{array}
\usepackage{subcaption}
\mathtoolsset{showonlyrefs}
\usepackage{multirow}
\usepackage{braket}
\usepackage{graphicx}      
\usepackage{natbib}        

\usepackage{algorithm}
\usepackage{algorithmic}

\usepackage{xcolor}

\newtheorem{ass}{Assumption}
\newcommand{\cmark}{\checkmark} 
\newcommand{\xmark}{\ding{55}}  

\newcolumntype{M}[1]{>{\centering\arraybackslash}m{#1}}

\usepackage{braket}
\begin{document}
\begin{frontmatter}
\title{On Frequency-Weighted Extended Balanced Truncation\thanksref{footnoteinfo}} 
%
\thanks[footnoteinfo]{This work is supported by the Swedish Research Council under the grants 2023-04770 and 2024-00185}
\author[P,KTH]{Sribalaji C. Anand} 
\author[KTH]{Henrik Sandberg} 
\address[P]{Department of Electrical and Systems Engineering, University of Pennsylvania, United States.}
\address[KTH]{
School of Electrical Engineering and Computer Science, KTH Royal Institute of Technology, Sweden (email: \{srca,hsan\}@kth.se).}
\begin{abstract}                
This paper addresses the problem of frequency-weighted extended balanced truncation for discrete and continuous-time linear time-invariant plants. We show that the frequency-weighted discrete-time plant admits block-diagonal solutions to both the Lyapunov inequality and its extended form. A recursive algorithm for extended balanced truncation is proposed, together with corresponding a-priori error bounds. Theoretical results are extended to continuous-time systems and validated through numerical examples.
\end{abstract}
\end{frontmatter}
%
\section{Introduction}
Balanced truncation is a model-order reduction method used for linear time-invariant (LTI) plants and controllers. Balanced truncation is performed by first generating a balanced realization (using Lyapunov equations and coordinate transformation), and then truncating or residualizing the states. In this paper, we consider the problem of frequency-weighted balanced truncation.

Frequency-weighted (FW) balanced truncation has received considerable research attention from the research community. The problem was first considered by \cite{enns1984model} where the author proposed a method to derive the reduced-order model. An error bound for two-sided weights was derived in \cite{kim1995error}, and the resulting bound is itself rather involved. It was also observed that the method proposed by \cite{enns1984model} {may} produce unstable reduced-order models in the presence of both input and output weights. Thus, the method was extended to produce stable reduced order model in the presence of two-sided weights \citep{wang2002new,imran2014frequency}.

While constructing the balanced realization, balancing the Gramians generally destroy internal model structures that may be important to preserve, such as the controller and plant states \citep{zhou1995structurally,beck2002model,li2005structured}. The structure is preserved by enforcing a block-diagonal structure on the solution of the Lyapunov inequalities (LIs)\citep{oh2002simple,li2003structured}. Generalized balanced truncation using these block diagonal solutions resulted in better error bound than \cite{kim1995error}. An overview of the available FW model reduction techniques can be found in \cite{sreeram2005frequency}, and \cite{astolff2024forty}.

It was shown by \cite{sandberg2010extension} that one can instead use the solutions of the extended LIs for balancing. {Extended LIs were introduced for control synthesis of DT plants by \cite{de2002extended}. Similarly, \cite{apkarian2001continuous} derived LIs with additional variables; these have also been used for controller synthesis, but not in the context of model reduction.} Extended balanced truncation often gives lower model reduction error and better error bounds. However, the extended LIs has not systematically been used to study the FW balanced truncation problem, which is the focus of this paper. {To this end, we consider an LTI plant with input and output weights. We then make the following contributions:}
\begin{enumerate}
    \item We prove that the FW discrete-time (DT) plant admits block-diagonal solutions to both the LI (Theorem~\ref{thm:1}) and the extended LI (Theorem~\ref{thm:2}).
    \item We formulate an optimization problem that minimizes the {upper bound of the worst-case} (infinity-norm) difference between the FW plant and the reduced plant. Since this problem is non-convex, we propose a relaxed recursive algorithm for extended balanced truncation (Algorithm~\ref{alg:EBT}) and provide the corresponding a-priori error bound (Theorem~\ref{thm:3}).
    \item We show that the error bound derived from {extended} balanced truncation is less than or equal to the bound from generalized balanced truncation (Proposition~\ref{prop:bound}).
    \item We develop an algorithm that applies extended balanced truncation to continuous-time (CT) plants and derive the associated error bound (Algorithm~\ref{alg:EBT:CT}). {To the best of our knowledge, such bounds were previously presented in \cite{borja2021extended} under the additional assumption that the slack variables are equal -- a restriction we do not impose.}
\end{enumerate}

\emph{Outline:} The remainder of the paper is organized as follows. We introduce the problem in Section~\ref{sec:PF}, and provide an overview of the standard approaches for model reduction. We provide the iterative algorithm for extended balanced truncation in Section~\ref{sec:EBT}, and its CT counterpart in Section~\ref{sec:discuss}. The effectiveness of the proposed method is depicted through numerical examples in Section~\ref{sec:NE}, and concluding remarks are presented in Section~\ref{sec:con}.

\emph{Notation:} Given a vector $x \in \mathbb{R}^n$, $\vert x \vert^2 := \sum_{i=1}^{n} x_i^2$. For any given real and proper transfer function $G$, $\Vert G \Vert_{\infty} := \sup_{\omega \in \mathbb{R}} \vert G(j\omega) \vert$. Let $A \in \mathbb{R}^{n \times n}$, then $\lambda_i(A), i \in \{1,\dots,n\}$ denotes the eigenvalues of $A$, and $\rho(A)= \max_{i} \vert \lambda_i(A)\vert$ denotes the spectral radius of $A$. A positive (semi-)definite matrix is denoted by $A\succ 0 \; (A \succeq 0)$. Let $M \in \mathbb{R}^{m \times n}$ be any real matrix, then $\sigma_i(M) := \sqrt{\lambda_i (M^TM)}$ denotes the $i$-th singular value of $M$, and $\bar{\sigma}(M) = \max_i \sigma_i(M)$ denotes the largest singular value. {The nuclear norm is denoted by $\|M\|_* := \sum_{i=1}^{\min(m,n)} \sigma_i(M)$.}
\section{Problem description}\label{sec:PF}
Consider a DT plant represented by 
\begin{equation}\label{eq:plant}
\begin{aligned}
x[k+1] &= Ax[k] + Bu[k], \\
y[k] &= Cx[k]+Du[k],
\end{aligned}
\end{equation}
where $x[k] \in \mathbb{R}^n$, $u[k] \in \mathbb{R}^m$, and $y[k] \in \mathbb{R}^p$. The corresponding transfer function is defined as 
\begin{equation}
G(z) := D+ C(zI - A)^{-1}B.
\end{equation}
Similarly, we define the input and output weights as 
\begin{align}
W_i(z) &:= D_i + C_i(zI - A_i)^{-1}B_i, \\
W_o(z) &:= D_o + C_o(zI - A_o)^{-1}B_o,
\end{align}
where the tuples $(A_i, B_i, C_i, D_i)$ and $(A_o, B_o, C_o, D_o)$ are the state-space realizations associated with $W_i(z)$ and $W_o(z)$, respectively. The state dimensions of the input and output weights are denoted by $n_i$ and $n_o$, respectively. {We next adopt the following assumption.}
\begin{ass}
{The matrices $A$, $A_i$ and $A_o$ are Schur stable, i.e., $\rho(A)<1$, $\rho(A_i)<1$, $\rho(A_o)<1$. $\hfill \triangleleft$}
\end{ass}

The objective of FW model order reduction is to find a reduced-order model $G_r(z) := D_r + C_r(zI - A_r)^{-1}B_r$ such that the quantity
\begin{equation}\label{eq:error:bound}
\gamma_e := \Vert W_o(G - G_r)W_i \Vert_{\infty}
\end{equation}
is minimized. Here, $r$ is the reduced model order with $r \ll n$. Next we outline some of the existing approaches in the literature to solve the problem. 
\subsection{Standard approaches to model order reduction}
%
{An algorithm for FW \emph{balanced truncation} was first proposed in CT \citep[Table~1]{enns1984model}.} However, there are two limitations of the algorithm. Firstly, the algorithm {may} produces unstable reduced order plants in the presence of input and output weights. Secondly, there are no explicit {a-priori} error bounds. While the quality of the approximation can be explicitly determined by calculating the value of \eqref{eq:error:bound}, there are no computationally efficient and straightforward methods to compute the error bound. However, both of the above shortcomings were addressed in the generalized balanced truncation approach \citep{zhou1995structurally} which we outline next.
%

Consider the FW plant $W_o(z)G(z)W_i(z)$, whose state-space representation is given by
\begin{equation}\label{eq:matrices}
\big[\tilde{A} \;|\; \tilde{B} \;|\; \tilde{C}^T\big] =
\left[
\begin{array}{ccc|c|c}
A_o & 0 & B_oC & 0 & C_o^T \\
0 & A_i & 0 & B_i & 0 \\
0 & BC_i & A & BD_i & (D_oC)^T
\end{array}
\right].
\end{equation}
Then the generalized FW balanced truncation is given in Algorithm~\ref{alg:BT}. There are two main observations from Algorithm~\ref{alg:BT}. Firstly, the error bound can be given by the simple expression \eqref{eq:bound:exp}. Secondly, to preserve the plant's state structure {under balanced transformation}, we enforce a block-diagonal structure on $\tilde{P}$. Next, we demonstrate the existence of such block-diagonal solutions, thereby presenting the first contribution of this paper. 

\begin{algorithm}[tb]
\caption{Generalized FW Balanced Truncation}
\label{alg:BT}
\begin{algorithmic}[1]
  \STATE Solve the Lyapunov ineqalities
        \begin{align}
        &\tilde{A}\tilde{P}\tilde{A}^T-\tilde{P}+\tilde{B}\tilde{B}^T \prec 0, \quad \tilde{P} \succ 0 \label{gram:P}\\
	&\tilde{A}^T\tilde{Q}\tilde{A}-\tilde{Q}+\tilde{C}\tilde{C}^T \prec 0, \quad \tilde{Q} \succ 0 \label{gram:Q}\\
        &\text{where}\; \tilde{P} = \mathrm{diag}(P_{oi}, P), \quad \tilde{Q} = \mathrm{diag}(Q_{oi}, Q), \label{gram:Pst}\\
     	&P_{oi}, Q_{oi} \in \mathbb{R}^{(n_o+n_i) \times (n_o+n_i)}, P, Q \in \mathbb{R}^{n \times n} \label{temp3}
        \end{align}
    \STATE Construct the state transformation matrix $S$:
    \begin{enumerate}
        \item Find a matrix $F$ such that $P = F^T F$.
        \item Diagonalize $FQF^T$ as $FQF^T = U Z^2 U^T$.
        \item Set $S = F^T U Z^{-1/2}$.
    \end{enumerate}
    \STATE Compute the error bound $\gamma_b$:
\begin{equation}\label{eq:bound:exp}
    \gamma_e \leq {2\sum_{i=r+1}^n \sqrt{\lambda_i(PQ)}} = 2 \sum_{i=r+1}^{n} \xi_i :=\gamma_b,
    \end{equation}
    where $\xi$ are the diagonal elements of $Z$ and $\gamma_e$ is the reduction error in \eqref{eq:error:bound}.
        \STATE Apply the coordinate transformation:
    \begin{equation}
     \bar{A} = S^{-1} {A} S, \; \bar{B} = S^{-1} {B}, \; \bar{C} = {C} S, \; \bar{D} = {D}.
    \end{equation}
    \STATE Pick $r$, truncate the realization $(\bar{A}, \bar{B}, \bar{C}, \bar{D})$ and construct $G_r(z)$.
\end{algorithmic}
\end{algorithm}
\begin{thm}\label{thm:1} 
There exist positive definite matrices \(\tilde{P} \succ 0\) and \(\tilde{Q} \succ 0\) that have the block structures in \eqref{gram:Pst}--\eqref{temp3} and satisfy the LIs \eqref{gram:P} and \eqref{gram:Q}. \(\hfill \square\)
\end{thm}
The proof of Theorem~\ref{thm:1}, and other results in the sequel, can be found in the appendix. Following the approach in \cite{oh2002simple}, the result of Theorem~\ref{thm:1} can likewise be extended to the case of single-sided weightings.

In general, the extended balanced truncation provides improved error and bounds compared to the above methods \citep{sandberg2010extension,borja2021extended}. {Thus, we next apply extended balanced truncation for FW model reduction.}
\begin{rem}\label{rem:trace}
An upper bound on the truncation error \(\gamma_e\) is given by the sum of the Hankel singular values. To minimize this bound, a good heuristic is to select matrices \(P\) and \(Q\) that satisfy \eqref{gram:P} and \eqref{gram:Q}, respectively, with the smallest trace {\citep{bendotti2002role}}. \(\hfill \triangleleft\)
\end{rem}
\section{Extended balanced truncation}\label{sec:EBT}
The central idea of extended balanced truncation is to use the extended LIs for constructing a balanced realization. Analogous to \eqref{gram:P} and \eqref{gram:Q}, the extended LIs for the weighted plant \eqref{eq:matrices} is given by
\begin{equation}\label{eq:ELE1} 
\begin{bmatrix}
\tilde{P} & \tilde{A}\tilde{R} & \tilde{B} \\
\tilde{R}^T \tilde{A}^T & \tilde{R} + \tilde{R}^T - \tilde{P} & 0 \\
\tilde{B}^T & 0 & I
\end{bmatrix} \succ 0,
\end{equation}
\begin{equation}\label{eq:ELE2} 
\begin{bmatrix}
\tilde{N} + \tilde{N}^T - \tilde{Q} & \tilde{N}\tilde{A} & 0 \\
\tilde{A}^T \tilde{N}^T & \tilde{Q} & \tilde{C}^T \\
0 & \tilde{C} & I
\end{bmatrix} \succ 0,
\end{equation}
%
%
%
where, the matrices \( \tilde{R} \) and \( \tilde{N} \) may be non-symmetric. In extended balanced truncation, the additional degrees of freedom introduced by the matrices \(\tilde{R}\) and \(\tilde{N}\) can be leveraged to further reduce the error bound. 

Next we show that there exists block diagonal solutions satisfying \eqref{eq:ELE1} and \eqref{eq:ELE2} marking a contribution of the paper.
\begin{thm}\label{thm:2}
There exist positive definite matrices \(\tilde{P} \succ 0\) and \(\tilde{Q} \succ 0\), and symmetric block-diagonal matrices 
\begin{equation}\label{grma:RNst}
\begin{aligned}
\tilde{R} = \tilde{R}^T &= \mathrm{diag}(R_{oi}, R),\;\;\;\tilde{N} = \tilde{N}^T = \mathrm{diag}(N_{oi}, N),\\
R_{oi}, N_{oi} &\in \mathbb{R}^{(n_o+n_i) \times (n_o+n_i)}, \;\;R, N \in \mathbb{R}^{n \times n},
\end{aligned}
\end{equation}
that satisfy the extended LIs \eqref{eq:ELE1} and \eqref{eq:ELE2}. \(\hfill \square\)
\end{thm}
We have shown that the extended LIs admit a block diagonal solution. We next show that when such block diagonal solution exists, then the error bound {on $\gamma_e$} is given by a simple formula. 
\begin{thm}\label{thm:3}
Suppose there exists positive definite matrices $\tilde{P} \succ 0, \tilde{Q} \succ 0,$ and symmetric block diagonal matrices \eqref{grma:RNst} such that \eqref{eq:ELE1} and \eqref{eq:ELE2} hold. Then there exists a coordinate transformation matrix $S$ such that
\begin{align}
S{R}S^T &= \left(S^{-1}\right)^T{N}S^{-1} = \Sigma = \mathrm{diag}\left( \Sigma_{1}, \Sigma_{2}\right) \label{ext:bal:gram}\\
\text{where}\;\Sigma_{1} &= \mathrm{diag}\left(\sigma_{1},\sigma_{2},\dots,\sigma_{r}\right),\\
\Sigma_{2} &= \mathrm{diag}\left(\sigma_{r+1},\sigma_{r+2},\dots,\sigma_{n} \right),
\end{align}
and $\sigma_{r+1} > \sigma_{r} \geq 0$. Let $G_r$ be the reduced order model obtained by truncating the $n-r$ states corresponding {to the realization $(S {A} S^{-1}, S {B},{C} S^{-1}, {D})$}.
Then it holds that 
\begin{equation}\label{ext:bound}
\Vert W_o(G - G_r)W_i \Vert_{\infty} \leq 2\; \mathrm{trace}(\Sigma_{2}). \qquad \square
\end{equation}
\end{thm}
We next develop an algorithm that minimizes the error bound in \eqref{ext:bound}. To this end, we rewrite \eqref{ext:bound} as 
\begin{equation}
2\; \mathrm{trace}(\Sigma_{2}) = 2 \sum_{i=r+1}^n \sqrt{\lambda_i(RN)} 
\end{equation}
which holds since the eigenvalues of the product $RN$ are invariant under coordinate transformation $S$ in \eqref{ext:bal:gram}. The error bound \eqref{ext:bound} is minimized $\forall \; r \in \{1,\dots,n\}$ when ${R}$ and ${N}$ are obtained by solving
\begin{equation}\label{eq:noncvx}
\begin{aligned}
\inf_{\tilde{R} = \tilde{R}^T,\;\tilde{N} =\tilde{N}^T} & \;  \sum_{i={1}}^n \sqrt{\lambda_i({R}{N})},\;\;\; \text{s.t.} \; \eqref{eq:ELE1},\eqref{eq:ELE2},\eqref{grma:RNst}.
\end{aligned}
\end{equation}
Although the constraints in \eqref{eq:noncvx} are linear matrix inequalities, the objective function is nonconvex. Consequently, \eqref{eq:noncvx} cannot be efficiently solved numerically. Therefore, we propose a convex suboptimal iterative balancing technique in Algorithm~\ref{alg:EBT}, marking our second contribution. Next, we provide a detailed discussion of the algorithm.
\begin{algorithm}[tb]
\caption{Iteration algorithm to solve the FW extended balanced truncation problem }
\label{alg:EBT}
\begin{algorithmic}[1]
  \STATE Obtain a block diagonal solution to \eqref{gram:P} and \eqref{gram:Q}. 
  \STATE Let $\bar{P}=P$, $\bar{Q}=Q$, and {set} loop\_count$=0$.\\ \textbf{while} loop\_count $\leq$ loop\_max \textbf{do}
   	\STATE \hspace{3em} Solve the convex optimization problem 
    		\begin{equation}\label{eq:nuc:P}
    		\begin{aligned}
    			\min_{\tilde{N},\tilde{Q}} & \quad \Vert \bar{P}N\Vert_{*}\\
    			\mathrm{s.t.} & \quad \eqref{eq:ELE2}, \eqref{grma:RNst}, N \preceq \bar{Q}
        		\end{aligned}
        \end{equation}
        	\STATE \hspace{3em} $\bar{Q} \gets N$.
        	\STATE \hspace{3em} Solve the convex optimization problem 
    		\begin{equation}\label{eq:nuc:Q}
   			\begin{aligned}
    				\min_{\tilde{R},\tilde{P}} & \quad \Vert R\bar{Q}\Vert_{*}\\
    				\mathrm{s.t.} & \quad \eqref{eq:ELE1}, \eqref{grma:RNst}, R \preceq \bar{P}
        			\end{aligned}
        		\end{equation}
        \STATE \hspace{3em} $\bar{P} \gets R$.
        \STATE \hspace{3em} loop\_count $\gets$ loop\_count + 1\\
 	\textbf{end while}
	\STATE Construct the state transformation matrix $S$:
    \begin{enumerate}
        \item Find a matrix $F$ such that $R = F^T F$.
        \item Diagonalize $FNF^T$ as $FNF^T = U \Sigma^2 U^T$.
        \item Set $S = F^T U \Sigma^{-1/2}$.
    \end{enumerate}
    \STATE Compute the error bound:
		\begin{equation}\label{eq:bound:exp2}
    			\gamma_{eb} := {2 \sum_{i=r+1}^{n} \sqrt{\lambda_i(RN)}} = 2 \sum_{i=r+1}^{n} \xi_i,
    		\end{equation}
    		where $\xi_i$ are the diagonal elements of $\Sigma$.
	\STATE Apply the coordinate transformation:
    		\begin{equation}
     			\bar{A} = S^{-1} {A} S, \; \bar{B} = S^{-1} {B}, \; \bar{C} = {C} S, \; \bar{D} = {D}, 
   	 	\end{equation}
    	\STATE Pick $r$, truncate the realization $(\bar{A}, \bar{B}, \bar{C}, \bar{D})$ and construct $G_r(z)$.
\end{algorithmic}
\end{algorithm}
\section{Discussion}\label{sec:discuss}
In this section, we discuss three aspects of the proposed algorithm. First, we explain how the optimization problems \eqref{eq:nuc:P} and \eqref{eq:nuc:Q} in Algorithm~\ref{alg:EBT} relate to the original non-convex problem in \eqref{eq:noncvx}. Second, we show that the error bound from Algorithm~\ref{alg:EBT} is lower or equal to the one from Algorithm~\ref{alg:BT}. Finally, we briefly discuss the continuous-time version of extended balanced truncation.
\subsection{Convex proxy for the nonconvex objective in \eqref{eq:noncvx}}
As previously mentioned, the objective function in \eqref{eq:noncvx} is nonconvex. Therefore, we aim to replace the objective with a convex proxy. To this end, we next reformulate the objective function in \eqref{eq:noncvx}.
\begin{prop}\label{lem:inequality}
Let there exist $\tilde{P} \succ 0$, $\tilde{Q} \succ 0$, $\tilde{R} = \tilde{R}^T$ and $\tilde{N}=\tilde{N}^T$ that satisfy \eqref{eq:ELE1}, \eqref{eq:ELE2}, and \eqref{grma:RNst}. Then
\begin{equation}\label{eq:bound}
\sum_{i=1}^n \sqrt{\lambda_i(RN)} \leq \sqrt{n \|RN\|_*} \qquad \hfill \square
\end{equation}
\end{prop}
Proposition~\ref{lem:inequality} shows that $\sqrt{n}\,\sqrt{\|RN\|_*}$ serves as an upper bound for the non-convex objective in \eqref{eq:noncvx} (see Remark~\ref{rem:tight} about the tightness of the upper bound). Since the factor $\sqrt{n}$ is independent of the optimization variables, it can be dropped. Moreover, because the square root is a monotonically increasing function, minimizing $\sqrt{\|RN\|_*}$ is equivalent to minimizing $\|RN\|_*$. Hence, we can use $\|RN\|_*$ as a proxy for the objective in \eqref{eq:noncvx}.  

Nevertheless, $\|RN\|_*$ remains non-convex (bi-linear) since both $R$ and $N$ are optimization variables. To address this, we employ an alternating minimization strategy: we fix $R$ and optimize over $N$, then fix $N$ and optimize over $R$, iterating this process; this iterative process along with the balancing procedure is presented in Algorithm~\ref{alg:EBT}. The constraints $N \preceq \bar{Q}$ and $R \preceq \bar{P}$ are added to improve the error bound at each iteration \citep{sandberg2012parameterized}. Next, we show that the error bound does not deteriorate while using Algorithm~\ref{alg:EBT}, marking our third contribution. 
\begin{prop}\label{prop:bound}
Let $P$ and $Q$ be the lower blocks of the block-diagonal solutions of the LMIs \eqref{gram:P} and \eqref{gram:Q}, respectively. Similarly, let $R$ and $N$ denote the lower blocks of the block-diagonal solutions of the extended LMIs \eqref{eq:ELE1} and \eqref{eq:ELE2}, respectively. Let us define
\[
{\gamma_b = 2\sum_{i=1}^n \sqrt{\lambda_i(PQ)}, \quad \gamma_{eb} = 2\sum_{i=1}^n \sqrt{\lambda_i(RN)}}.
\]  
Then it holds that $\gamma_{eb} \leq \gamma_b. \hfill \square$
\end{prop}
Since the error bound for (extended) balanced truncation can be expressed as the sum of the singular values of the product of the Gramians, it follows from Proposition~\ref{prop:bound} that when we solve Algorithm~\ref{alg:EBT}, which involves solving the extended LMIs \eqref{eq:ELE1}--\eqref{eq:ELE2}, we can be assured that the error bound does not deteriorate. We later demonstrate this through numerical examples. However, achieving the improved error bound comes at the cost of increased computational complexity due {to} larger LMIs.
\subsection{Continuous-time Extended balanced truncation}
%
%
More recently, a detailed study on CT extended balanced truncation was conducted by \cite{borja2021extended}. Although the problem is addressed in detail, their error bound is derived under certain equality assumptions. For instance, in their Theorem~3, the bound is obtained when $\alpha = \beta$. In contrast, our work presents a model reduction method that does not rely on any such assumptions.

To this end, let us denote the CT transfer functions as
\begin{align}
{ G(s)} &:= D_c + C_c (sI - A_c)^{-1} B_c\\
{ W_i(s)} &:= D_{i,c} + C_{i,c} (sI - A_{i,c})^{-1} B_{i,c}\\
{ W_o(s)} &:= D_{o,c} + C_{o,c} (sI - A_{o,c})^{-1} B_{o,c}
\end{align}
Similar to \eqref{eq:matrices}, let us denote the state-space matrices of the weighted CT plant as $(\tilde{A}_c, \tilde{B}_c,\tilde{C_c})$. Next we introduce Algorithm~\ref{alg:EBT:CT} for extended balanced truncation in CT. 

\begin{algorithm}[tb]
\caption{Iteration algorithm to solve the FW extended balanced truncation problem in CT}
\label{alg:EBT:CT}
\begin{algorithmic}[1]
  \STATE Obtain a block diagonal solution to \eqref{gram:P} and \eqref{gram:Q}. 
  \STATE Pick $t \in \mathbb{R}_{>0}$, let $\bar{P}=P$, $\bar{Q}=Q$, and $\kappa=\frac{t}{2}$.\\ \textbf{while} loop\_count $\leq$ loop\_max \textbf{do}
   \STATE \hspace{3em} Solve the {convex} optimization problem 
    \begin{equation}\label{eq:nuc:P}
    \begin{aligned}
    \min_{\tilde{N},\tilde{Q}} & \quad \Vert \bar{P}N\Vert_{*}\\
    \mathrm{s.t.} & \quad \eqref{eq:ELE:CT1}, \eqref{grma:RNst}, N \preceq \bar{Q}
        \end{aligned}
        \end{equation}
        \STATE \hspace{3em} $\bar{Q} \gets N$.
        \STATE \hspace{3em} Solve the {convex} optimization problem 
    \begin{equation}\label{eq:nuc:Q}
	\begin{aligned}
    	\min_{\tilde{R},\tilde{P}} & \quad \Vert R\bar{Q}\Vert_{*}\\
    	\mathrm{s.t.} & \quad \eqref{eq:ELE:CT2},\eqref{grma:RNst}, R \preceq \bar{P}
        \end{aligned}
        \end{equation}
        \STATE \hspace{3em} $\bar{P} \gets R$.
        \STATE \hspace{3em} loop\_count $\gets$ loop\_count + 1\\
 	\textbf{end while}
                \STATE Construct the matrix $S$ (step 8 in Algorithm \ref{alg:EBT})
    \STATE Apply coordinate transformation to the plant:
    \begin{align}
     \bar{A} &= S^{-1} \left( I-\kappa{A}_c \right)^{-1}\left( I+\kappa{A}_c \right)S,\\
      \bar{B} &= S^{-1} \left( I-\kappa{A_c}\right)^{-1}{B}_c{t},\; \bar{C} = {C}_c \left( I-\kappa{A}_c \right)^{-1} S,\\
      \bar{D} &= {{D}_c + \kappa C_c \left( I-\kappa{A}_c \right)^{-1}B_c}. 
    \end{align}
    \STATE Compute the error bound:
\begin{equation}\label{eq:bound:exp3}
    \Vert {W}_o \left( {G}_c - {G}_{c,r}\right){W}_o\Vert _{\infty}:= 2 \sum_{i=r+1}^{n} \xi_i,
    \end{equation}
    where $\xi_i$ are the Hankel singular values.
	\STATE Pick $r$, truncate the realization $(\bar{A}, \bar{B}, \bar{C}, \bar{D})$ and construct $G_{d,r}(z):=D_{d,r} + C_{d,r}(zI-A_{d,r} )^{-1}B_{d,r} $
 	\STATE Derive ${G}_{c,r} = D_{c,r} + C_{c,r}(zI-A_{c,r} )^{-1}B_{c,r} $ where
        		\begin{align}
     		{A}_{c,r} &= \frac{1}{\kappa}\left( A_{d,r} -I \right)\left( {A}_{d,r}  + I \right)^{-1},\\
     		{B}_{c,r} &= \frac{1}{t} \left( I-\kappa A_{c,r} \right)B_{d,r}, \;{C}_{c,r} =  {C}_{d,r} \left( I-\kappa {A}_{c,r}  \right),\\
	{D}_{c,r}  &= {{D}_{d,r} - \kappa C_{c,r} \left( I-\kappa{A}_{c,r} \right)^{-1}B_{c,r}}
    		\end{align}

\end{algorithmic}
\end{algorithm}

\begin{figure*}[htbp]
    \centering
    \begin{minipage}{0.49\textwidth}
        \begin{equation}\label{eq:ELE:CT1}
\begin{bmatrix}
N_c+N_c^T-Q_c & N_c \left( 1+\kappa{A}_c\right) & 0\\
\star & \left( 1-\kappa{A}_c\right)^T Q_c  \left( 1-\kappa{A}_c\right) & {C}_c^T\\
\star & \star &I
\end{bmatrix} \succ 0
        \end{equation}
    \end{minipage}
    \hfill
    \begin{minipage}{0.49\textwidth}
        \begin{equation}\label{eq:ELE:CT2}
           \; \begin{bmatrix}
\left( 1-\kappa{A}_c\right) P_c  \left( 1-\kappa{A}_c\right)^T & \left( 1+\kappa{A}_c\right) R_c & {B}_c\\
\star & R_c+R_c^T+P_c & 0\\
\star & \star & \frac{1}{t^2}I
\end{bmatrix} \succ 0
        \end{equation}
    \end{minipage}
\begin{align}\label{eq:1}
&\begin{bmatrix}
P_{oi,1}-A_o P_{oi,1} A_o^T - B_o C P C^T B_o^T & P_{oi,2}-A_o P_{oi,2} A_i^T & -A_o P_{oi,2} C_i^T B^T - B_o C P A^T \\
\ast & P_{oi,3}- A_i P_{oi,3} A_i^T -B_i B_i^T & - A_i P_{oi,3} C_i^T B^T- B_i D_i^T B^T \\
\ast & \ast & L_{33}
\end{bmatrix} \succ 0,\\
&L_{33} := P - A P A^T- B C_i P_{oi,3} C_i^T B^T- B D_i D_i^T B^T.
\end{align}
    \noindent\rule{\textwidth}{0.4pt}  
\end{figure*}

The overall idea behind Algorithm~\ref{alg:EBT:CT} is as follows. Consider the FW CT plant, which is discretized using a bilinear (Tustin) transformation. The variable $t$ in Algorithm~\ref{alg:EBT:CT} represents the sampling time. Steps~$1$ through $7$ are similar to those in Algorithm~\ref{alg:EBT}; however, in this case, the extended LI are derived for the discretized weighted plant. For numerical well-posedness and to avoid matrix inversions, congruence transformations are applied to the extended LIs, resulting in \eqref{eq:ELE:CT1} and \eqref{eq:ELE:CT2} (see Appendix~\ref{app:congruence} for the proof). Once the Gramians are obtained from the extended LIs, the coordinate transformation matrix is constructed (step~$8$), applied to the DT plant (step~$9$), truncated (step~$11$), and then converted back to CT (step~$12$).

The advantages of the algorithm are threefold. Firstly, Algorithm~\ref{alg:EBT:CT} is one of the first in the literature to provide an error bound for CT plants with weighting in a simple form (step~$10$). {This error bound arises from the property that, for CT systems, the bilinear (Tustin) transformation preserves the infinity norm of the weighted plant \citep{khalil1996robust} since it is a bijection between the imaginary axis (CT) and the unit circle (DT)}. As a result, discretization preserves the infinity norm, and an error bound derived for the DT weighted plant serves as an error bound for the CT plant as well.  

Secondly, there is no need for explicit discretization of the weighted plant. {In general, for a given CT plant, earlier studies suggested} discretizing the weighted plant, performing extended truncation on the DT system, and then converting the reduced plant back to CT \citep{sandberg2008model}. By contrast, Algorithm~\ref{alg:EBT:CT} avoids this explicit discretization step. We also note that although matrix inverses of the plant are formally involved in Steps~$9$ through $11$ for (re-)discretization, they can be avoided in practice by using factorization methods such as \emph{linsolve} in MATLAB, improving both numerical stability and efficiency.

\begin{rem}\label{rem:CTDT}
{Although discretization is not necessary to run Algorithm~\ref{alg:EBT:CT}, the first line of the algorithm assumes a block-diagonal solution of the DT LI corresponding to the DT matrices. Nevertheless, similar techniques based on congruence transformations, as applied to the extended LI in Appendix~\ref{app:congruence}, can be employed to avoid discretization. $\hfill \triangleleft$}
\end{rem}

Finally, since the variable $t$ appears explicitly in the extended LIs \eqref{eq:ELE:CT1} and \eqref{eq:ELE:CT2}, it can be adjusted to obtain tighter error bounds. The choice of $t$ affects the discretization step size and selecting a smaller or larger $t$ can (possibly) improve the conservativeness of the bound depending on the system dynamics. {We also note that, in comparison to \cite{borja2021extended}, where there are two slack variables ($\alpha$ and $\beta$), our method has only one slack variable $t$.}

Moreover, if algorithms for solving nonlinear optimization problems are available, the variable $t$ can be treated as a decision variable and optimized to minimize the derived error bound, subject to the constraint that all required matrix inverses remain well-defined. This flexibility allows Algorithm~\ref{alg:EBT:CT} not only to provide guaranteed error bounds but also to exploit the sampling time as a tuning parameter to achieve the best possible approximation quality for a given CT plant.

\section{Numerical Examples}\label{sec:NE}
\begin{figure}
    \centering
    \includegraphics[width=8cm]{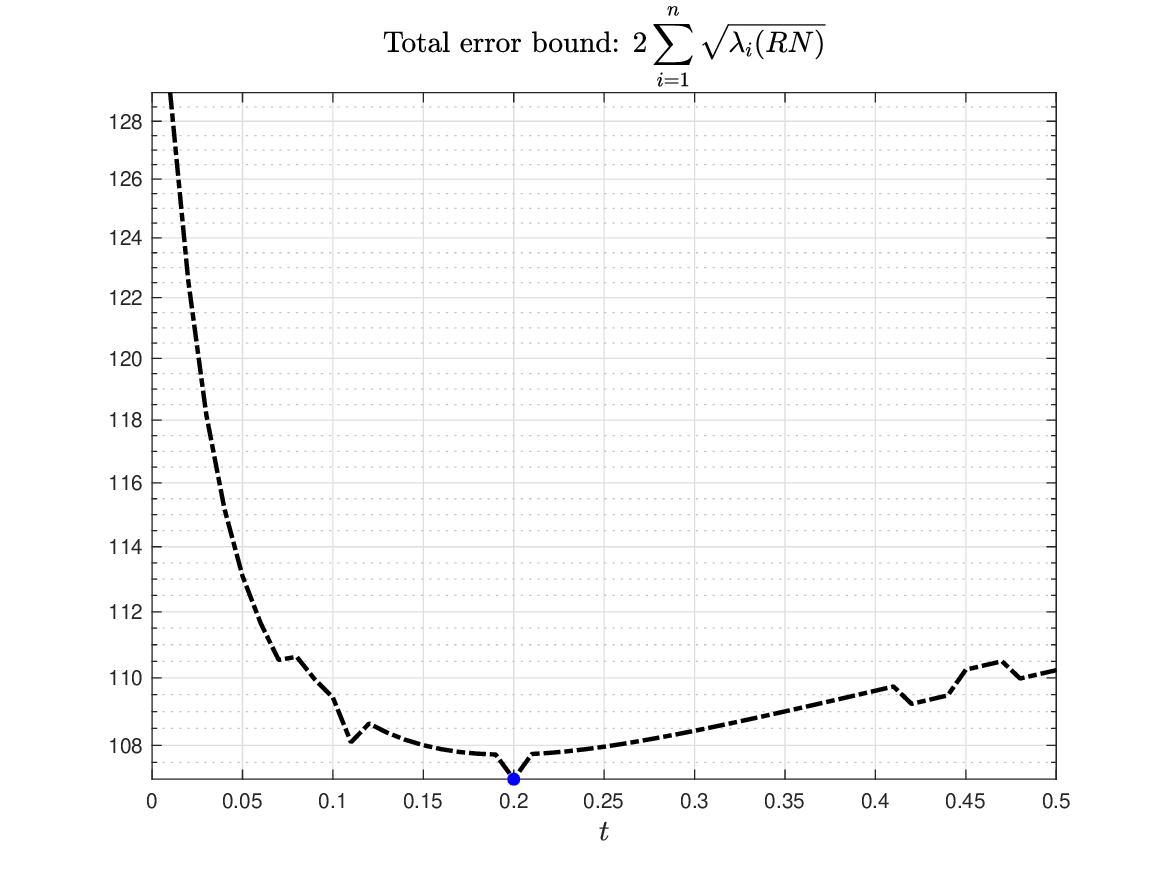}
    \caption{Comparison of error bound for varying values of $t$.}
    \label{fig:error}
\end{figure}

\begin{figure}
    \centering
    \includegraphics[width=8cm]{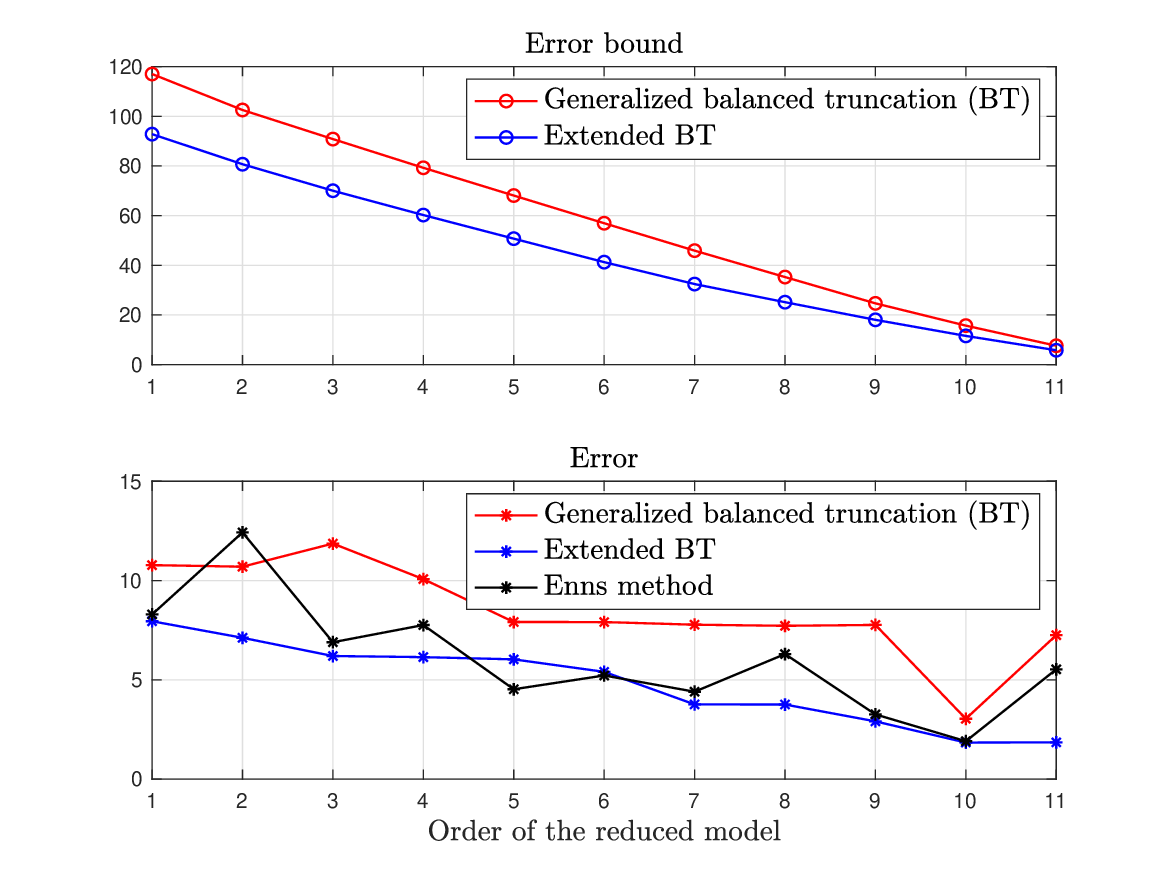}
    \caption{Comparison of errors and bounds for Example~\ref{exmp:1}.}
    \label{plot:ex:1}
\end{figure}

\begin{figure}
    \centering
    \includegraphics[width=8cm]{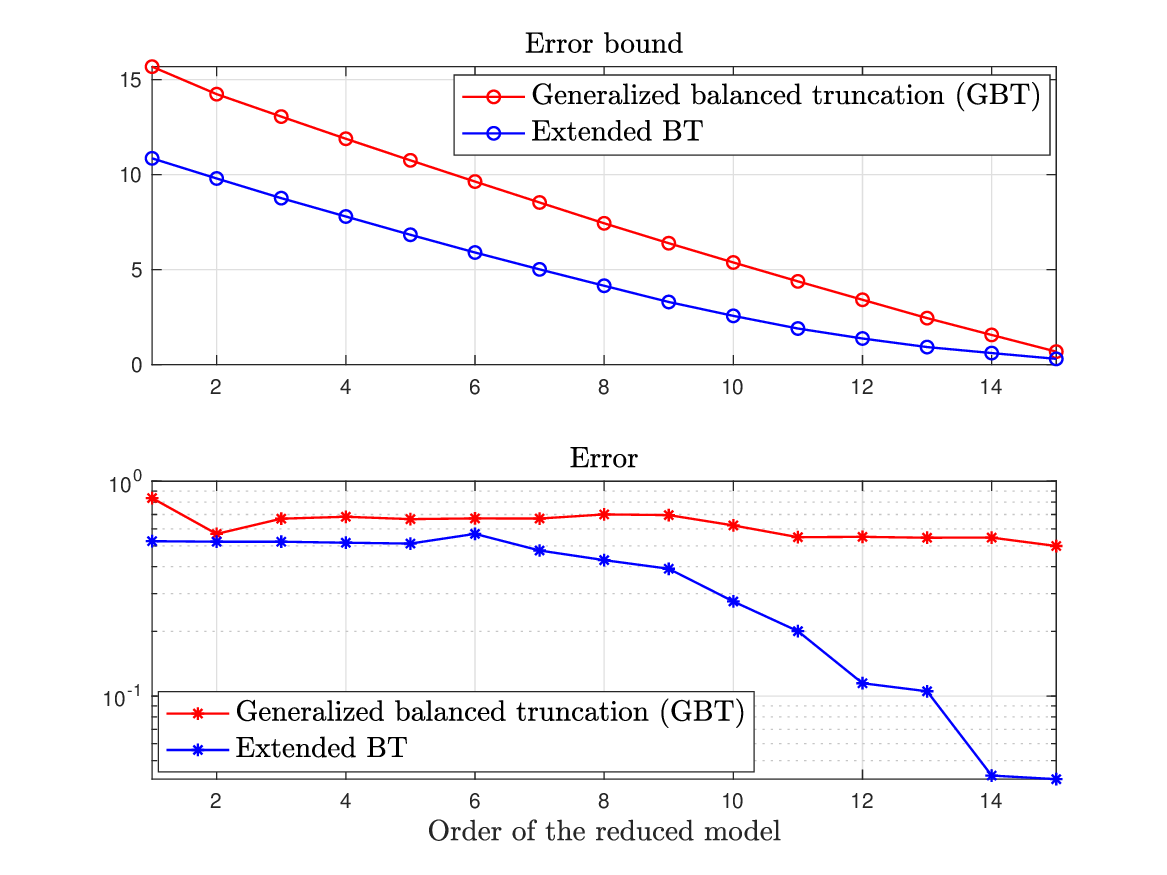}
    \caption{Comparison of errors and bounds for Example~\ref{exmp:2}.}
    \label{plot:ex:2}
\end{figure}

\begin{figure}
    \centering
    \includegraphics[width=8cm]{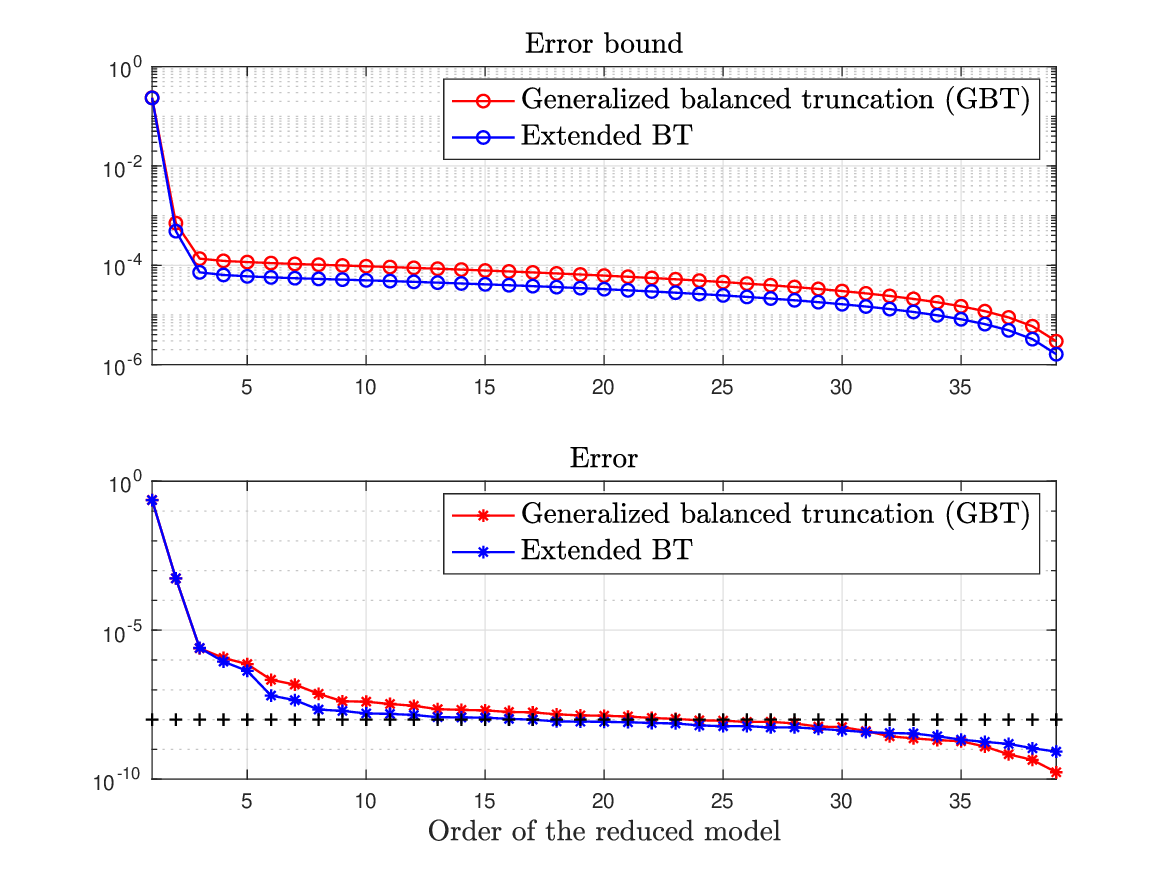}
    \caption{Comparison of errors and bounds for Example~\ref{exmp:3}.}
    \label{plot:ex:3}
\end{figure}

In this section, we depict the efficacy of the proposed algorithms via numerical examples. An overview of the examples is given in Table.~\ref{tab:comparison}. The numerical examples are solved in MATLAB 2024a using YALMIP \citep{lofberg2004yalmip} and MOSEK 11 \citep{aps2019mosek}.
\begin{table*}[h!]
\centering
\caption{Overview of the numerical examples.}
\label{tab:comparison}
\begin{tabular}{|c|M{0.9cm}|M{0.7cm}|M{0.77cm}|c|c|M{1.7cm}|M{1cm}|M{4.4cm}|c|}
\hline
{Example} & Plant, Weights & Input weight & Output weight & $n$ & {Poles} & {Disc. Method} & Reduced plant & {Comparison} & Result \\ \hline
$1$ & CT & \xmark  & \cmark & 12 & Resonant & None (implicitly Tustin)  & CT & Enns method \citep{enns1984model}, and CT generalized balanced truncation \citep{oh2002simple}  & Fig.~\ref{plot:ex:1}\\ \hline
$2$ & CT &\cmark & \cmark & 16 & Resonant & Tustin  & DT &DT Balanced truncation (Algorithm \ref{alg:BT}) & Fig.~\ref{plot:ex:2} \\ \hline
$3$ & CT &  \cmark & \xmark & 50 & Real     & FOH    & DT & DT Balanced truncation (Algorithm \ref{alg:BT}) & Fig.~\ref{plot:ex:3}\\ \hline
\end{tabular}
\end{table*}
\begin{exmp}\label{exmp:1}
We consider a resonant $12^{\text{th}}$-order system
\begin{equation}\label{eq:num:plant}
{G}(s) = \sum_{k=1}^{6} \frac{\omega_k^2}{s^2 + 2 {\zeta} \omega_k s + \omega_k^2},
\end{equation}
where \(\omega_k \in \{1, 2, 10, 20, 35, 50\}\) and \({\zeta} = 0.1\). The frequency weight \(W_o(s)\) is a band-pass filter given by
${W}_o(s) = \frac{s}{(s/5 + 1)^2}.$
First, we derive the reduced-order models using Enn's method\footnote{Since the weighting is applied only to the outputs, Enn's method is guaranteed to produce a stable model.} and the balanced truncation method of \cite{oh2002simple}, which is the CT counterpart of Algorithm~\ref{alg:BT}. The Gramians in Algorithm~\ref{alg:BT} are obtained by minimizing the trace (see Remark~\ref{rem:trace}). {The model reduction errors are shown in Fig.~\ref{plot:ex:1} (bottom). Since Enn's method does not provide an a priori error bound, we show only the error bounds obtained from \cite{oh2002simple} in Fig.~\ref{plot:ex:1} (top).}

Next, we derive the reduced-order CT model using extended balanced truncation. {To this end, we apply Algorithm~\ref{alg:EBT:CT} to obtain the reduced CT plant together with its corresponding error bound. The algorithm is executed for various values of $t$, and we select the value that yields the smallest bound $2 \sum_{i=1}^n \sqrt{\lambda_i(RN)}$. The values of the error bound for different choices of $t$ are shown in Fig.~\ref{fig:error}. Since $t = 0.2$ achieves the smallest bound (highlighted by the blue dot in Fig.~\ref{fig:error}), we plot the associated error and error-bound curves (for varying $r$) in Fig.~\ref{plot:ex:1}.}  


{There are three main observations from Fig.~\ref{plot:ex:1}. First, the error bounds are quite conservative for both approaches. Second, our algorithm consistently outperforms Enn's method and achieves tighter bounds than balanced truncation. Finally, the model reduction error decreases monotonically with respect to the model order only for our method.} \hfill$\square$
\end{exmp}
\begin{exmp}\label{exmp:2}
We consider a resonant $16^{\text{th}}$-order system
\begin{equation}\label{eq:num:plant}
{G}(s) = \sum_{k=1}^{8} \frac{\omega_k^2}{s^2 + 2 {\zeta} \omega_k s + \omega_k^2},
\end{equation}
where \(\omega_k \in \{1, 2, 3, 5, 10, 20, 50, 80\}\) and \({\zeta} = 0.1\). The frequency weights \(W_o(s)\) and \(W_i(s)\) are a band-pass and low-pass filter, respectively, given by
\begin{equation}\label{eq:num:weights}
{W}_o(s) = \frac{s}{(s/5 + 1)^2}, \;\text{and}\;{W}_i(s) = \frac{1}{s + 10}.
\end{equation}
We discretize the transfer functions using \emph{bilinear} sampling with a sampling period of \(0.1 \mbox{s}\). Let us denote the discretized transfer functions as \(G(z), W_o(z)\) and \(W_i(z)\). We obtain the reduced order models using balanced truncation in Algorithm~\ref{alg:BT}, and extended balanced truncation in Algorithm~\ref{alg:EBT}. The results are shown in Fig.~\ref{plot:ex:2} and depict that our algorithm outperforms Algorithm~\ref{alg:BT}. $\hfill \square$
\end{exmp}
\begin{exmp}\label{exmp:3}
We consider a single-input single-output plant, where the system matrices are randomly generated \citep{willcox2002balanced}. Specifically, the matrix $A_c \in \mathbb{R}^{n \times n}$ is an upper-triangular matrix with $n=40$, diagonal elements $A_{c,ii} \in [-0.1,\;-0.15]$ and off-diagonal elements in the interval $[0,\;0.001]$. The input and output matrices, $B_{c} \in \mathbb{R}^{n \times 1}$ and $C \in \mathbb{R}^{1 \times 20}$, are sampled element-wise from the interval $\left[0,\;1\right]$. The input weight is given by $
W_{i,c}(s) = \frac{1}{10s+1}$
The transfer functions are discretized using \emph{first-order hold (FOH)} with a sampling period of \(0.1 \mbox{s}\). Let us denote the discretized transfer functions as \(G(z)\) and \(W_i(z)\), respectively. We obtain the reduced order models using balanced truncation in Algorithm~\ref{alg:BT}, and extended balanced truncation in Algorithm~\ref{alg:EBT}. 

The results are depicted in Fig.~\ref{plot:ex:3} which shows the mean values of the error bound and the model reduction error, computed over $N = 10$ randomized trials. It can be observed from the plot that our method yields a tighter error bound, and a stringent error bound. The error computed by Algorithm~\ref{alg:BT} is less than the error obtained by our Algorithm~\ref{alg:EBT}. However, the discrepancy is approximately $10^{-8}$, which is equivalent to the tolerance of the solver (MOSEK) employed. Consequently, the errors can be considered negligible. $\hfill \square$
\end{exmp}
\section{Conclusions}\label{sec:con}
In this paper, we addressed the problem of FW extended balanced truncation for LTI plants. We proposed a recursive algorithm for extended balanced truncation applicable to both CT and DT systems. The effectiveness of the proposed method was demonstrated through numerical examples and comparisons with existing approaches. As shown in Fig.~\ref{plot:ex:1}, the resulting error bound is fairly conservative. Understanding the source of this conservatism is an interesting direction for future work.
%
\bibliography{ifacconf}             
\appendix
\section{Proof of Theorem~\ref{thm:1}}
We provide the proof only for the reachability Gramian since the proof for the observability Gramian is similar.
\begin{equation}\label{pf:2}
\text{Let}\; P_{oi} = \begin{bmatrix} P_{oi,1} & P_{oi,2}\\ \ast & P_{oi,3}
\end{bmatrix}.
\end{equation}
Substituting \eqref{pf:2}, \eqref{eq:matrices} and \eqref{gram:Pst} into \eqref{gram:P}, we obtain \eqref{eq:1}. Let us denote the block elements of \eqref{eq:1} as \( L_{ij} \).

Since the matrix \( A_i \) is Schur, there exists a matrix \( P_{oi,3} \succ 0 \) such that
\[
P_{oi,3} - A_i P_{oi,3} A_i^T - Q \succ 0
\]
for any given \( Q \succeq 0 \). Let \( Q = B_i B_i^T \succeq 0 \). Then it follows that there exists \( P_{oi,3} \succ 0 \) such that \( L_{22} \succ 0 \) holds. Let this be Observation 1 (O1).

Consider the matrix $\bar{L} := \begin{bmatrix} L_{22} & L_{23} \\ \ast & L_{33} \end{bmatrix}.$
We aim to show that \( \bar{L} \succ 0 \). Since \( L_{22} \succ 0 \) (from O1), it follows from the property of the Schur complement \cite{boyd1994linear} that
\begin{equation}\label{pf:1}
\bar{L} \succ 0 \iff \bar{L} / L_{22} := L_{33} - L_{23}^T L_{22}^{-1} L_{23} \succ 0.
\end{equation}
Using the definition of the matrix $L_{33}$, we can rewrite \eqref{pf:1} as $\bar{L} \succ 0$ holds iff 
\begin{equation}\label{eq:pf0}
P - A P A^T -Q \succ 0
\end{equation}
where $Q:= B C_i P_{oi,3} C_i^T B^T + B D_i D_i^T B^T + L_{23}^T L_{22}^{-1} L_{23} \succeq 0$. Now, since the matrix \( A \) is Schur, there exists a matrix \( P \succ 0 \) such that \eqref{eq:pf0} holds. Thus it holds that $\bar{L} \succ 0$. Let this be Observation 2 (O2).

Next, we aim to show that \( L \succ 0 \). Since we know that \( \bar{L} \succ 0 \) (from O2), it follows that
$L \succ 0 \iff L / \bar{L} \succ 0.$x
Using the definition of the matrix $L_{11}$, we can rewrite the condition as $L \succ 0$ holds iff 
\begin{equation}\label{eq:pf1}
P_{oi,1} - A_o P_{oi,1} A_o^T -Q \succ 0
\end{equation}
where $Q:= B_o C P C^T B_o^T + \begin{bmatrix} L_{12} & L_{13} \end{bmatrix} \bar{L}^{-1} \begin{bmatrix} L_{12}^T \\ L_{13}^T \end{bmatrix}\succeq 0$. Now, since the matrix \( A_o \) is Schur, there exists a matrix \( P_{oi,1} \succ 0 \) such that \eqref{eq:pf1} holds. Thus it holds that $\exists \tilde{P}$ such that $\bar{L} \succ 0$ which concludes the proof. 
%
%
\section{Proof of Theorem~\ref{thm:2}}
We prove the result for the reachability Gramian; the proof for the observability Gramian follows analogously. 

From the statement of the theorem, since the matrices \( A_o, A_i, \) and \( A \) are Schur stable, we know from Theorem~\ref{thm:1} that there exists a positive definite block-diagonal matrix $\tilde{P}$ that satisfies \eqref{gram:P}. Then we have:
\begin{equation}
\begin{aligned}
\begin{bmatrix}
\tilde{P} - \tilde{A} \tilde{P} \tilde{A}^T - \tilde{B} \tilde{B}^T & 0 \\
0 & \tilde{P}
\end{bmatrix} \succ 0 
&\iff 
\begin{bmatrix}
\tilde{P} - \tilde{B} \tilde{B}^T & \tilde{A} \tilde{P} \\
\tilde{P} \tilde{A}^T & \tilde{P}
\end{bmatrix} \succ 0 \\
&\iff 
\begin{bmatrix}
\tilde{P} & \tilde{A} \tilde{P} & \tilde{B} \\
\tilde{P} \tilde{A}^T & \tilde{P} & 0 \\
\tilde{B}^T & 0 & I
\end{bmatrix} \succ 0,
\end{aligned}
\end{equation}
using the Schur complement \citep{boyd1994linear}. This shows that inequality \eqref{eq:ELE1} is satisfied by setting \( \tilde{R} = \tilde{R}^T = \tilde{P} \), which concludes the proof.
\section{Proof of Theorem~\ref{thm:3}}
Suppose $\tilde{R}$ and $\tilde{N}$ are symmetric solutions of the extended LIs \eqref{eq:ELE1} and \eqref{eq:ELE2}. Then it follows that $\tilde{R} \succ 0$ and $\tilde{N} \succ 0$, and consequently, $R \succ 0$ and $N \succ 0$. By using Theorem 7.5 in \cite{zhou1998essentials}, there exists a matrix $S$ such that \eqref{ext:bal:gram} holds. Furthermore, since $R_{oi} \succ 0$ and $N_{oi} \succ 0$, there also exists a matrix $S_{oi}$ such that
\[
S_{oi}R_{oi}S_{oi}^T = (S_{oi}^{-1})^TN_{oi}{S_{oi}}^{-1} = \Sigma_{oi}
\]
with $\Sigma_{oi} = \mathrm{diag}(\sigma_{oi,1}, \sigma_{oi,2}, \dots, \sigma_{oi,n_o+n_i})$. 

Let us define \(\tilde{S}= \text{diag}(S_{oi}, S)\), \(\tilde{\Sigma} = \text{diag}(\Sigma_{oi,}, \Sigma)\). The state-space matrices corresponding to the transfer function $W_o(z)G(z)W_i(z)$ in the transformed coordinate is given as
\[
\left[\begin{array}{ccc|c}
S_{oi}A_{11}S_{oi}^{-1} & S_{oi} \begin{bmatrix}{B_o\hat{C}}\\0
\end{bmatrix} & S_{oi} \begin{bmatrix}B_o\hat{C}_{12}\\0
\end{bmatrix} & S_{oi}B_1\\[6pt]
\begin{bmatrix}
0 & \hat{B}C_i
\end{bmatrix}S_{oi}^{-1}  & \hat{A} & \hat{A}_{12} & \hat{B}D_i\\
\begin{bmatrix}
0 & \hat{B}_{12}C_i
\end{bmatrix}S_{oi}^{-1} & \hat{A}_{21}& \hat{A}_{22} & \hat{B}_{12}D_i\\ \hline 
C_1S_{oi}^{-1} & D_o\hat{C} & D_o\hat{C}_{12} & 0
\end{array}\right],
\]
\begin{align}
\text{where} \;
\left[
\begin{array}{c|c}
A_{11} & B_{1} \\ \hline
C_{1} & 0
\end{array}
\right] = \left[ 
\begin{array}{cc|c}
A_o & 0 & 0 \\
0 & A_i & B_i \\ \hline
C_o & 0 & 0
\end{array}
\right]
\end{align}
The state-space matrices corresponding to $W_o(z)G(z)W_i(z)$ has balanced extended controllability and observability gramians $\left(\tilde{S}\tilde{P}\tilde{S}^T,\Sigma \right)$ and $\left( (\tilde{S}^{-1})^T\tilde{Q}\tilde{S}^{-1},\Sigma \right)$. 

The state-space matrices corresponding to the reduced-order transfer function $W_o(z)G_r(z)W_i(z)$ is given by 
\[
\left[\begin{array}{cc|c}
S_{oi}A_{11}S_{oi}^{-1} & S_{oi} \begin{bmatrix}{B_o\hat{C}}\\0
\end{bmatrix}  & S_{oi}B_1\\[6pt]
\begin{bmatrix}
0 & \hat{B}C_i
\end{bmatrix}S_{oi}^{-1}  & \hat{A}  & \hat{B}D_i\\ \hline 
C_1S_{oi}^{-1} & D_o\hat{C} &  0
\end{array}\right].
\]
Since the $W_o(z)G_r(z)W_i(z)$ is the extended balanced truncation of $W_o(z)G(z)W_i(z)$, the proof follows from Theorem~2 in \cite{sandberg2010extension}.
\section{Proof of Proposition~\ref{lem:inequality}}
Suppose $\tilde{R}$ and $\tilde{N}$ are symmetric solutions of the extended LIs \eqref{eq:ELE1} and \eqref{eq:ELE2}.  
From these inequalities, it follows that $\tilde{R} \succ 0$ and $\tilde{N} \succ 0$, and consequently $R \succ 0$ and $N \succ 0$.

Since \(R\) and \(N\) are symmetric positive definite, \(R^{1/2}\) is well defined and $S := R^{1/2} N R^{1/2}$ is symmetric positive definite. Hence \(S\) has positive real eigenvalues, and \(S\) is similar to \(RN\):
\[
RN = R^{1/2} \big(R^{1/2} N R^{1/2}\big) R^{-1/2} = R^{1/2} S R^{-1/2} \sim S,
\]
so \(RN\) and \(S\) share the same spectrum. In particular, \(\lambda_i(RN) > 0\) for every \(i\).

A standard inequality for any square matrix \(A\) is
\begin{equation}\label{eq:lambda_sigma}
\sum_{i=1}^n |\lambda_i(A)| \le \sum_{i=1}^n \sigma_i(A) = \|A\|_*,
\end{equation}
Letting \(A = RN\) in \eqref{eq:lambda_sigma} and using \(\lambda_i(RN) > 0\) yields
\begin{equation}\label{eq:pf:1}
\sum_{i=1}^n \lambda_i(RN) \le \|RN\|_*.
\end{equation}
The Cauchy--Schwarz inequality states that for any vector \(a \in \mathbb{R}^n\), $\Big|\sum_{i=1}^n a_i\Big|^2 \le n \sum_{i=1}^n |a_i|^2$. Applying this to the vector \(\big(\sqrt{\lambda_1(RN)}, \ldots, \sqrt{\lambda_n(RN)}\big)\) gives
\[
\left( \sum_{i=1}^n \sqrt{\lambda_i(RN)} \right)^2
\le n \sum_{i=1}^n \lambda_i(RN)
\le n \|RN\|_*,
\]
where the last inequality follows from \eqref{eq:pf:1}. Taking square roots establishes \eqref{eq:bound}, which concludes the proof.
\begin{rem}\label{rem:tight}
The Cauchy--Schwarz inequality is tight exactly when the entries \(\sqrt{\lambda_i(RN)}\) are all equal, i.e., when all eigenvalues of \(RN\) are equal to the same positive constant \(c\), so \(RN = cI\). The intermediate inequality \(\sum_i \lambda_i(RN) \le \|RN\|_*\) is an equality precisely when the multiset of singular values \(\{\sigma_i(RN)\}\) equals the multiset of absolute eigenvalues \(\{|\lambda_i(RN)|\}\); in particular, this holds when \(RN\) is symmetric positive semidefinite. Therefore, the inequality \eqref{eq:bound} holds with equality if and only if \(RN = cI\) for some \(c > 0\).
\end{rem}

\section{Proof of Proposition~\ref{prop:bound}}
Assume, for contradiction, that $\gamma_{eb} > \gamma_b$. Let $\tilde{P}$ and $\tilde{Q}$ be the solution of the LMIs \eqref{gram:P} and \eqref{gram:Q}, respectively. By construction, $(\tilde{P},\tilde{R}=\tilde{P})$ and $(\tilde{Q},\tilde{N}=\tilde{Q})$ are also feasible for the extended LMIs \eqref{eq:ELE1} and \eqref{eq:ELE2}, respectively.  But then there exists a feasible solution of the extended LMIs with singular-value sum $\gamma_b$, which contradicts our assumption that the optimal extended solution has $\gamma_{eb} > \gamma_b$. Hence, the assumption is false, and we conclude that $\gamma_{eb} \le \gamma_b$. which concludes the proof.

\section{Derivation of (\ref{eq:ELE:CT1}) and (\ref{eq:ELE:CT2})}\label{app:congruence}
We provide the proof only for the reachability Gramian \eqref{eq:ELE:CT2} since the proof for the observability Gramian \eqref{eq:ELE:CT1} is similar. Consider the CT system $(\tilde{A}_c,\tilde{B}_c,\tilde{C}_c)$ which is dicretized using bilinear transformation and the system matrices are given by $(\tilde{A},\tilde{B}, \tilde{C}):=$
\begin{equation}\label{eq:matrices:pf}
\left((I-\kappa \tilde{A}_c)^{-1}(I+\kappa \tilde{A}_c),(I-\kappa \tilde{A}_c)^{-1}\tilde{B}_ct, \tilde{C}_c(I-\kappa \tilde{A}_c)^{-1} \right)
\end{equation}
Consider the extended LI \eqref{eq:ELE1} which can be written as 
\begin{equation}\label{pf:e1}
\begin{bmatrix}
\tilde{P}-\tilde{B}\tilde{B}^T & \tilde{A}\tilde{R} \\
\tilde{R}^T \tilde{A}^T & \tilde{R} + \tilde{R}^T - \tilde{P}
\end{bmatrix}\succ 0
\end{equation}
using Schur Complement. Substiting the matrices \eqref{eq:matrices:pf} in \eqref{pf:e1}, and applying congruence transformation with respect to the matrix $\begin{bmatrix}
I-\kappa \tilde{A}_c & 0\\ 0 & I
\end{bmatrix}
$. Then taking Schur complement with respect to the top left block yields the desired LMI. 
\end{document}